# Real-Time Hybrid Simulation for Infrastructure Degradation Assessment: Conceptual Framework and Illustrative Application

Manuel Salmeron[1,*], Herta Montoya[2], Edwin Patino[3], Ingrid Madera Sierra[4] and Shirley Dyke[5]

[1] *Lyles School of Civil Engineering, Purdue University, West Lafayette, IN*
[2] *Escuela de Arquitectura, Universidad del Valle, Cali, Colombia*
[3] *School of Mechanical Engineering, Purdue University, West Lafayette, IN*
* Corresponding author: *salmeron@purdue.edu*

**ABSTRACT:** To date, the prospect of using real-time hybrid simulation (RTHS) to study the effects of long-term or "wear-and-tear" loads, such as exposure to harmful environmental conditions or fatigue, has remained underexplored. This study presents a conceptual framework to assess the impact of long-term degradation on infrastructure systems. The framework integrates the capabilities of RTHS with accelerated degradation techniques to evaluate the behavior of a degrading system over time. Experimental results obtained in this way not only capture the complex interactions but also provide a reliability-based method to determine the expected time-to-failure of the evaluated system. The developed framework is demonstrated using a virtual RTHS platform designed to test fiber-reinforced elastomeric isolators.

## 1. INTRODUCTION

Real-time hybrid simulation (RTHS) is a novel experimental technique that divides the studied system into numerical and physical subsystems interacting through actuation and sensor systems. This approach maintains the integrity of a system's interactions, unlike traditional testing methods that may oversimplify or isolate behaviors. The versatility of RTHS has made it suitable to evaluate different systems under various loads, such as base excitation (Condori Uribe et al., 2023; Najafi et al., 2023), wind loads (Al-Subaihawi et al., 2020), and extreme thermal loads for space habitats (Montoya et al., 2023).

Current applications of RTHS focus on assessing the effects that transient or short-term loads, such as earthquakes and wind, have on infrastructure systems (Palacio-Betancur and Gutierrez-Soto, 2023). However, the effects of long-term or "wear-and-tear" loads, such as exposure to harmful environmental conditions or fatigue, have remained underexplored due to the prohibitive long duration needed for these phenomena. This work presents a conceptual framework to assess the impact of long-term degradation on infrastructure systems. The framework leverages the partitioned nature of RTHS by subjecting the physical specimen to different degradation levels through accelerated degradation techniques.

## 2. FRAMEWORK OVERVIEW

One of the main benefits of RTHS experiments, compared to traditional testing or purely numerical simulation, is its ability to capture the systemic behavior of the studied system. It does so by introducing the realistic response of an experimental part while enforcing its interaction with a numerical simulation. The criteria to choose the physical subsystem or part often depend on the (1) research focus (what behavior do we want to investigate?), (2) experimental feasibility (what can we build in the laboratory?), or (3) economic feasibility (what is more cost-effective to build for testing?). In this framework, we propose introducing "degradability" as an additional criterion, defined as the possibility of observing changes due to degradation of the physical subsystem *within the time frame of the experimental tests*. If the chosen physical subsystem has this property, a series of short-term-impact (e.g., earthquakes or wind loads) RTHS experiments can be executed at regular intervals spanning a long enough time frame to assess the impact of the degraded physical subsystem on the overall system's performance.

Degradation processes often occur over a long time, especially aging phenomena. Hence, their effects on the tested specimen may not be noticeable until months or sometimes years later. Although keeping an RTHS testbed active for a long time is technically possible, it would reduce the flexibility of such a testbed for other experiments, dismissing one of RTHS' most valuable advantages. We propose using accelerated degradation (AD) techniques on the physical subsystem to overcome this difficulty. AD has been widely used to assess the isolated performance of various devices, ranging from mechanical to electronic (McPherson, 2019). Integrating AD techniques with RTHS would allow the examination of performance with degraded elements while capturing their interaction when coupled with other subsystems.

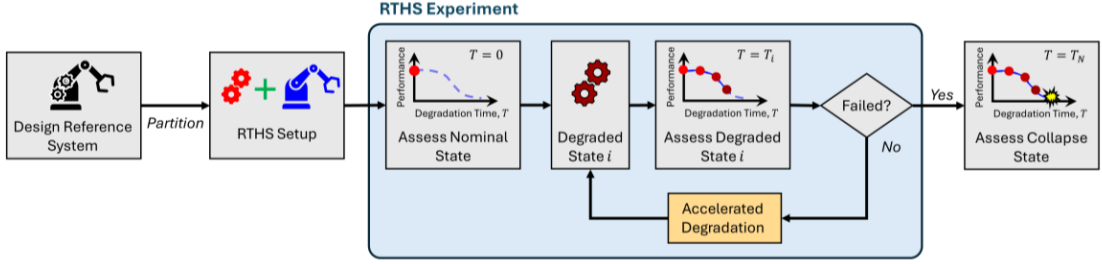

*Figure 1. Workflow describing the stages of the developed framework for RTHS of long-term degrading systems.*

The framework for long-term-impact RTHS is summarized in Fig. 1. First, a nominal-state RTHS test is conducted to assess the non-degraded performance of the system. The physical specimen is then subjected to AD, which shifts it to the next degraded state. At each degradation state, short-term-load RTHS experiments are conducted to assess the behavior of the degraded system. If a previously defined failure threshold has not been reached, the physical specimen is submitted to AD again, and a new set of short-term-load RTHS experiments is conducted. Once the system meets the established failure criteria, the time-to-failure is recorded based on the level of degradation reached during the last AD process.

## 3. APPLICATION

The framework described in the previous section is demonstrated by evaluating the long-term performance of a 3-story building with base isolation subjected to ground acceleration. Rubber isolators are used at the base to reduce the building's acceleration and drift. However, these affordable base isolators experience long-term degradation due to high relative humidity, which is known to reduce the flexibility of rubber over time (Gu et al., 2005). The RTHS partitioning for the experiment is shown in Fig. 2(a). AD is performed on the physical specimens by subjecting them to increased relative humidity conditions in a climatic chamber. After a trade-off study, the performance metrics of the degraded system are defined as appearing on Table 1. A testbed was built at the Intelligent Infrastructure Systems Laboratory (IISL) at Purdue University to test scaled rubber isolators. Consider the lumped-parameter model of the studied system in Fig. 2(b). Hereafter, a subscript $n$ denotes a variable belonging to the numerical subsystem, while a subscript $e$ denotes a physical subsystem variable.

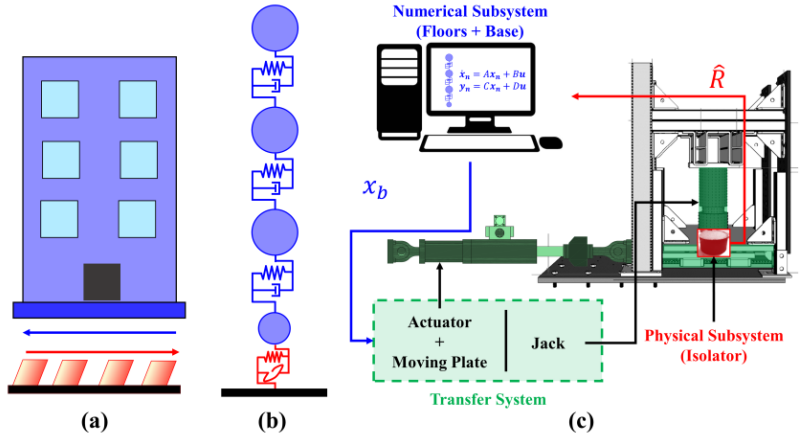

*Figure 2. RTHS for rubber isolators: (a) partitioned system; (b) lumped-parameter system; (c) experimental setup.*

Defining a state variable $Z_n = [X_n \quad \dot{X}_n]^T$ and the input as $u = [\ddot{x}_g \quad F_e]^T$, the partitioned dynamic equations of the system in state-space are:

$$\dot{Z}_n = A_n Z_n + B_n u \tag{1}$$

where $X_n = [x_b \quad x_1 \quad x_2 \quad x_3]^T$ is a vector containing the relative displacement of the base, $x_b$, and of the three stories, $x_j, j = 1,2,3$; $\ddot{x}_g$ is the ground acceleration; and $F_e = [R \quad 0 \quad 0]^T$, where $R$ is the measured restoring nonlinear force of the experimental seismic isolator. The matrices $A_n, B_n, C_n$ and $D_n$ are defined in terms of the system's dynamic properties.

The coupling between the numerical and physical subsystems is enforced through the transfer system shown in Fig. 2(c). A vertical hydraulic jack applies the load of the simulated building's weight (20 $kN$). The RTHS workflow is as follows: the numerical subsystem simulation is executed in the target machine, which sends the base displacement for the horizontal hydraulic actuator, with a moving plate attached, to implement it on the physical subsystem. In general, the realized and measured physical displacement, $x_m$, is not the same as the desired $x_b$. Thus, a controller between the transfer and physical systems is necessary to reduce the tracking error, $e = x_b - x_m$. A load cell measures the horizontal force needed to realize the desired displacement, including the moving plate's inertia. The force measurement and estimation strategy are used to estimate the restoring force, $\hat{R}$, of the isolator, which is sent back to the numerical subsystem and assembled as part of the feedback command $\boldsymbol{F}_e$.

A virtual realization of the RTHS testbed was used for this study. The control plant (i.e., the transfer system and the physical subsystem) models of this virtual RTHS platform are identified using experimental results from a series of open-loop experiments to evaluate the performance of low-cost rubber isolators under severe weather conditions. The transfer system, containing the hydraulic actuator and moving plate, is modeled as a parametric dynamic system (Maghareh et al., 2018). The restoring force of the seismic isolator is modeled using the Bouc-Wen model of hysteresis found in Song and Dyke (2013):

$$R = kx_b + \alpha z \tag{2}$$

where $k$ and $\alpha$ are stiffness variables, and $z$ is the hysteretic variable.

We consider that excessive relative humidity reduces the flexibility of rubber isolators over time (Polukoshko et al., 2018), for the simulation of the degradation process. For the virtual RTHS testbed, the degradation is modeled to follow a power law rule (McPherson, 2019):

$$k(T) = k_0[1 + A_0 T^m], \tag{3}$$

where $T$ is the exposure time to degrading conditions, $k_0 = 4.60 \times 10^4 \, N$ is the identified stiffness at nominal conditions (without degradation), $A_0$ is the material degradation coefficient, and $m$ is the time-dependence exponent. The increasing-stiffness hypothesis was confirmed by the open-loop experiments on the weather-degraded isolators: the horizontal stiffness of the exposed specimens increased by 2% and 12% when exposed to 14 and 31 days of increased relative humidity, respectively. Using this data, parameters $A_0$ and $m$ were fitted to $3.39 \times 10^{-5}$ and 2.37, respectively. Note that the isolator's behavior and degradation are modeled for the virtual RTHS platform. However, its behavior and response will be directly measured in the experimental realization of the RTHS framework.

Despite the nonlinear response of the scaled isolator, the dominant linear dynamics of the horizontal actuator and the moving table make the overall response predominantly linear. A PI controller with a lead-lag feedforward compensation reduces the tracking error ($e = x_b - x_m$) to less than 1.2%, which meets standard acceptance criteria in RTHS (Silva et al., 2020).

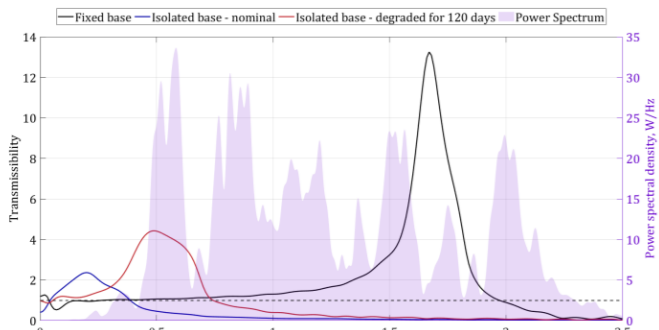

*Figure 3.* Comparison of the transmissibility between the fixed-base structure with the nominal and degraded seismically isolated systems.

Following the process described in Fig. 1, once the RTHS setup is completed, an RTHS test is performed to obtain the nominal state performance metrics. Transmissibility, defined as the ratio of maximum transmitted acceleration between the first floor and the ground, is a common metric to assess the efficacy of an isolating device (Chopra, 2007). Ideally, the isolation system should be less than 1 in the frequency regions where the ground motion is the most intense. Fig. 3 compares the transmissibility of the equivalent fixed-base building with the isolated building at nominal and after 120 days of AD. The power spectral density of the input excitation is shown on the secondary axis. Note that, when in its nominal state, the isolation successfully scales down the building's peak response and shifts it to the lower frequencies, avoiding the intense frequency range of the earthquake. After 120 days of exposure to AD, the same isolation system shows a higher peak response. Even more concerning is the intense phase of the earthquake when the transmissibility of the degraded isolation system is above 1, implying a response magnification. A comparison of the performance metrics at nominal state and after 120 days of AD is shown in Table 1. Even if the drift and acceleration requirements are met after degradation, the shear base and displacement requirements are exceeded.

*Table 1. Comparison of performance metrics between nominal and degraded-state systems. Violated metrics appear in red font.*

| Metric | Required Performance | Nominal State | 120 days of degradation |
|---|---|---|---|
| Interstory drift | $< 0.80\%$ | $0.003\%$ | $0.018\%$ |
| Acceleration on top | $< 2\,g$ | $0.08\,g$ | $0.38\,g$ |
| Shear base | $< 65\,kN$ | $23.43\,kN$ | $126.20\,kN$ |
| Maximum displacement | $< 14\,cm$ | $23.33\,cm$ | $35.48\,cm$ |
| Maximum base displacement | $< 35\,cm$ | $23.32\,cm$ | $35.34\,cm$ |

Now, suppose that more than one specimen is available for testing at each degraded state. In such a case, short-term-impact RTHS experiments can be done on each specimen. Thus, we define the time-to-failure ($TF$) of a single specimen as the exposure time at which at least one of the performance requirements is violated. If there are $N$ available devices, a probability distribution for $TF$ and associated statistics can be obtained. The parameters $k_0$, $A_0$, and $m$ of the degradation model of Eq. (4) can be considered stochastic to account for variations in the used materials and the fabrication process. Herein, we assume the number of available specimens is $N = 12$ and that the degradation model parameters are lognormally distributed with a mean equal to their nominal values and a coefficient of variation equal to 10%, 5%, and 5%, respectively. The resulting time-to-failure data is shown in the x-axis of Fig. 4(a) on a logarithmic scale. The data is fitted to a Weibull distribution, a common choice for time-to-degradation data (Feinberg, 2016; McPherson, 2019). The fitted scale parameter is 7.35, and the fitted shape parameter is 102.74. The Weibull distribution provides a good fit to the data at the 5% significance level. The fitted cumulative distribution function, or fragility curve, is shown in Fig. 4(b). The mean time-to-failure under AD conditions, $MTTF_{acc}$, is 96.35 days of exposure.

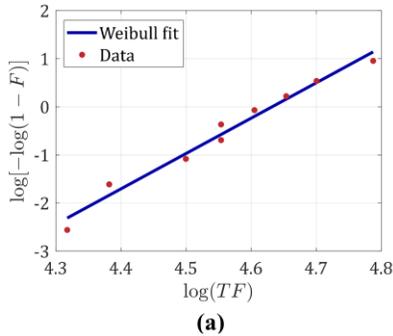
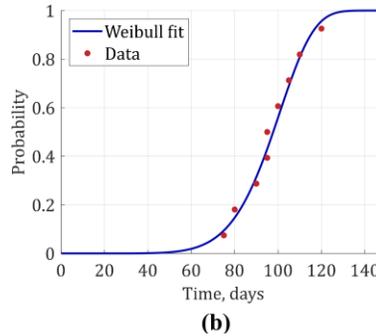

*Figure 4. Data and fitted Weibull distribution: (a) logarithmic scale; (b) cumulative distribution function.*

## 4. CONCLUSIONS

This work presents a framework to use RTHS for assessing the impact of long-term degradation on a system's performance. The framework combines the capability of RTHS to study system-level interactions with the ability of AD for examining slow degradation processes. The resulting tests yield insight on the performance loss with time that a system may have due to aging or increased operational conditions. Moreover, the partitioned nature of the experiment allows us to perform tests on multiple specimens, if available. Thus, reliability engineering techniques can be used to better diagnose the future performance of the tested system. An application example using rubber seismic isolators was shown. The virtual tests reveal a loss in isolation capacity of the specimens at 120 days, which is reflected in a transmissibility factor above 1 in the dominant frequencies of the used earthquake. Finally, a statistical analysis for a batch of isolators with simulated manufacturing uncertainty is shown. The process of generating a fragility curve and other helpful statistics using RTHS results is illustrated.

## ACKNOWLEDGEMENTS

This work was supported by the Space Technology Research Institutes grant (no. 80NSSC19K1076) from the National Aeronautics and Space Administration's Space Technology Research Grants Program.

# REFERENCES

Al-Subaihawi S, Kolay C, Marullo T, Ricles JM, Quiel SE. (2020). "Assessment of wind-induced vibration mitigation in a tall building with damped outriggers using real-time hybrid simulations." *Engineering Structures*, 205, 110044.

Chopra AK. (2007). *Dynamics of structures*. Pearson Education India.

Condori Uribe JW, Salmeron M, Patino E, Montoya H, Dyke SJ, Silva CE, ...,Montoya A (2023). "Experimental benchmark control problem for multi-axial real-time hybrid simulation." *Frontiers in Built Environment*, 9, 1270996.

Feinberg, A. (2016). *Thermodynamic Degradation Science: Physics of Failure, Accelerated Testing, Fatigue, and Reliability Applications*. John Wiley & Sons.

Gu H, Itoh Y, Satoh K. (2005). "Effect of rubber bearing ageing on seismic response of base-isolated steel bridges." In *Fourth International Conference on Advances in Steel Structures* (pp. 1627-1632). Elsevier Science Ltd.

Maghareh A, Silva CE, Dyke SJ. (2018). "Parametric model of servo-hydraulic actuator coupled with a nonlinear system: Experimental validation." *Mechanical Systems and Signal Processing*, 104, 663-672.

McPherson JW. (2019). "Reliability physics and engineering. Time-To-Failure Modeling, Third Edition" New York: Springer.

Montoya, H, Dyke, SJ, Silva, CE, Maghareh, A, Park, J, Ziviani, D .(2023). "Thermomechanical Real-Time Hybrid Simulation: Conceptual Framework and Control Requirements." *AIAA Journal*, 61(6), 2627-2639.

Najafi, A, Fermandois, GA, Dyke, SJ, and Spencer Jr, BF. (2023). "Hybrid simulation with multiple actuators: A state-of-the-art review." *Engineering Structures*, 276, 115284.

Palacio-Betancur, A, Gutierrez Soto, M. (2023). "Recent advances in computational methodologies for real-time hybrid simulation of engineering structures." Archives of Computational Methods in Engineering, 30(3), 1637-1662.

Pelliciari M, Marano GC, Cuoghi T, Briseghella B, Lavorato D, Tarantino AM. (2018). "Parameter identification of degrading and pinched hysteretic systems using a modified Bouc–Wen model." *Structure and Infrastructure Engineering*, 14(12), 1573-1585.

Polukoshko S, Martinovs A, Zaicevs E. (2018). "Influence of rubber ageing on damping capacity of rubber vibration absorber." *Vibroengineering Procedia*, 19, 103-109.

Silva CE, Gomez D, Maghareh A, Dyke SJ, Spencer Jr, BF. (2020). "Benchmark control problem for real-time hybrid simulation." *Mechanical Systems and Signal Processing*, 135, 106381.

Song W, Dyke S. (2014). "Real-time dynamic model updating of a hysteretic structural system." *Journal of Structural Engineering*, 140(3), 04013082.